\documentclass[a4paper]{article}

\usepackage{amsmath,amsthm,amsfonts,bbm}
\usepackage{longtable,multirow,multicol,wrapfig}
\usepackage{graphicx,float,stfloats,subcaption,caption}
\usepackage{algorithmicx,algorithm,diagbox}
\usepackage[compact]{titlesec}
\usepackage{tikz,booktabs,geometry,placeins}
\usetikzlibrary{shapes.geometric, arrows}
\usepackage[bottom]{footmisc}
\graphicspath{{plots/}}
\usepackage{setspace} 

\makeatletter
\let\@fnsymbol\@arabic
\AtBeginDocument{%
	\expandafter\renewcommand\expandafter\subsection\expandafter
	{\expandafter\@fb@secFB\subsection}%
	\newcommand\@fb@secFB{\FloatBarrier
		\gdef\@fb@afterHHook{\@fb@topbarrier \gdef\@fb@afterHHook{}}}%
	\g@addto@macro\@afterheading{\@fb@afterHHook}%
	\gdef\@fb@afterHHook{}%
}
\makeatother

\DeclareFontFamily{OMX}{yhex}{}
\DeclareFontShape{OMX}{yhex}{m}{n}{<->yhcmex10}{}
\DeclareSymbolFont{yhlargesymbols}{OMX}{yhex}{m}{n}

\usepackage[textsize=tiny, textwidth=1.5cm]{todonotes}

\newtheorem{theorem}{Theorem}

\newtheorem{ex}{Example}

\tikzstyle{startstop} = [rectangle, rounded corners, minimum width=4cm, minimum height=1cm,text centered, draw=black, fill=red!20]
\tikzstyle{io_impression} = [trapezium, trapezium left angle=70, trapezium right angle=110, minimum width=6cm, minimum height=1cm, text centered, text width=4cm, draw=black, fill=magenta!20]
\tikzstyle{io_attribution} = [trapezium, trapezium left angle=70, trapezium right angle=110, minimum width=2cm, minimum height=1cm, text centered, text width=2cm, draw=black, fill=green!20]
\tikzstyle{process} = [rectangle, minimum width=6cm, minimum height=1cm, text centered, text width=4cm, draw=black, fill=orange!10]
\tikzstyle{process_organic} = [rectangle, minimum width=4cm, minimum height=1cm, text centered, text width=4cm, draw=black, fill=blue!10]
\tikzstyle{decision} = [diamond, minimum width=2cm, minimum height=1cm, text centered, text width=2cm, draw=black, fill=green!10]
\tikzstyle{arrow} = [thick,->,>=stealth]
\tikzstyle{arrowdots} = [dotted,->,>=stealth]

\title{Show me the Money: Measuring Marketing Performance in Free-to-Play Games using Apple's App Tracking Transparency Framework}

 \author{Frederick Ayala-G\'omez\thanks{BANDAI NAMCO Mobile, fayala@bn-mobile.com; Torre Mapfre, Barcelona, Spain (Work was done at Rovio)(\textbf{Corresponding author})},
 	Ismo Horppu\thanks{Zynga, ihorppu@zynga.com; Aleksanterinkatu 9 A, Helsinki, Finland (Work was done while at Rovio)}, \\
 	Erlin Gülbenko\u{g}lu\thanks{Rovio Entertainment, \{\textit{name}\}.\{\textit{surname}\}@rovio.com; Keilaranta 7, Espoo, Finland}, 
 	Vesa Siivola$^{3}$, 
 	and Bal{\'a}zs Pej{\'o}\thanks{CrySyS Lab, pejo@crysys.hu; Magyar tudósok körútja 2, Budapest, Hungary}}

\date{}

\begin{document}

\maketitle

\begin{abstract}
	Mobile app developers use paid advertising campaigns to acquire new users. Marketing managers decide where to spend and how much to spend based on the campaigns' performance.
	Apple's new privacy mechanisms have a profound impact on how performance marketing is measured.
	Starting iOS 14.5, all apps must get system permission for tracking explicitly via the new App Tracking Transparency Framework, which shows the users a pop-up asking if they give the app permission to track.
	If a user does not allow tracking, the required identifier to deterministically find the online advertising campaign that brought the user to install the app is not shared.
	The lack of an identifier for attribution affects how the campaigns' performance is measured, as the users who do not allow tracking are not mapped one-to-one to a campaign.
	Instead of relying on individual identifiers, Apple proposed a new performance mechanism called conversion value, which is an integer set by the apps for each user, and the developers can get the number of installs per conversion value for each campaign.
	However, interpreting how conversion values are used to measure the campaigns performance is not obvious because it requires a method to translate the conversion values to revenue.
	This paper investigates the task of attributing revenue to advertising campaigns using the reported conversion values per campaign.
	Our contributions are to formalize the problem, find the theoretically optimal revenue attribution function for any conversion value schema, and show empirical results on past data of a free-to-play mobile game using different conversion value schemas.
\end{abstract}


\section{Introduction}

Mobile applications grow their player base profitably by acquiring players using paid advertising. The objective is to spend less than the revenue generated from the players to use the user acquisition budget in the best possible way.
Online advertising overcomes some of the problems highlighted by the famous quote from John Wanamaker: {\it ``Half the money I spend on advertising is wasted; the trouble is, I don’t know which half''}.
The quote refers to traditional advertisement channels such as newspapers, billboards, television, radio, leaflets, or any printed media, where the advertiser does not know which users engaged with an ad.
With the widespread usage of the internet and mobile devices, a new type of advertising media was born -- the so-called online media, where the marketing managers can measure the performance of the ad campaigns.
Online advertising includes, e.g., social networks, search engines, and ad networks, which benefit from knowing how users engage with the ads.
Besides the ability to track users who engaged with an ad, the internet and smart devices disrupted how companies advertise their products and services by allowing targeting of ads~\cite{blattberg1991interactive,shankar2009mobile,shankar2016mobile,winer2009new}.

Marketers look for online advertising channels that deliver the best return on investment (ROI).
Measuring ROI requires calculating the revenue that the ads campaign brought compared to the money spent.
For example, if a company invests \$100~USD in advertising an application and the users acquired through it bring \$200~USD in revenue the campaign is profitable with an ROI of 200\%.
Of course, the calculation of the return on investment requires attributing the user's revenue to the specific campaigns that brought them to the app.
This task is known as attribution, and there are different approaches to it~\cite{hughes2014multiple}.
The most common attribution model in online advertising is last-click attribution, which gives all the credit to the last ad that the user engaged with~\cite{dalessandro2012causally}.

In the search for higher ROI, online advertising companies started building user profiles.
Advertisers benefit from user profiles because they help them find more suitable ads, which increases the chances of converting from impression to action (e.g., click, purchase, install, subscribe).
For example, search engines show ads based on user-specific queries,
social networks promote products based on user interests,
and mobile apps show an advertisement based on the user's collected data.
User profiles enable new forms of advertising optimizations where companies may target users with specific criteria.
On the other hand, the more collected information there is in the user profiles, the more users become attentive to what the companies know about them.
Several surveys show that people are concerned about the control that companies have over their data, and they disagree with the data collection and sharing practices of online services~\cite{eurobarometer_eprivacy,us_ntia}.
Governments have taken action to rule how companies use personal data, which led to significant legislative changes.
Two recent and notable examples are the European General Data Protection Regulation~\cite{regulation2016regulation},
and the California Consumer Privacy Act of 2018~\cite{ccpa}.

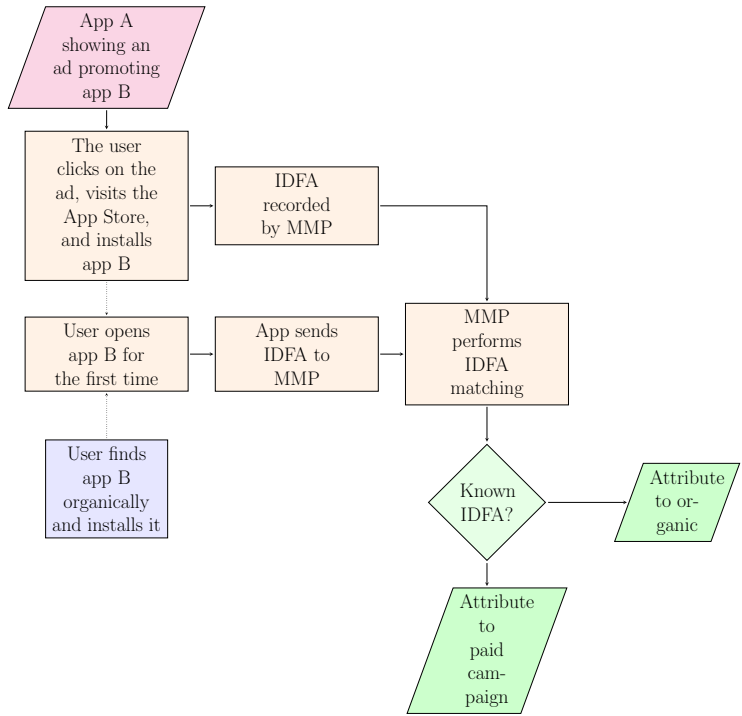
\begin{figure*}[t!]
	\centering
	\resizebox{.61\textwidth}{!}{
		\begin{tikzpicture}[node distance=5cm, inner sep=7, outer sep=2]
			\huge
			\node (io_imp) [io_impression] {App A showing an ad promoting app B};
			\node (pro_click) [process, below of=io_imp, yshift=-0.5cm] {The user clicks on the ad, visits the App Store, and installs app B};
			\node (pro_clickatt) [process, right of=pro_click, xshift=2cm] {IDFA recorded by MMP};

			\node (pro_openapp) [process, below of=pro_click, yshift=-0.5cm] {User opens app B for the first time};
			\node (pro_openappatt) [process, right of=pro_openapp, xshift=2cm] {App sends IDFA to MMP};

			\node (pro_organic) [process_organic, below of=pro_openapp] {User finds app B organically and installs it};

			\node (pro_match) [process, right of=pro_openappatt, xshift=2cm] {MMP performs IDFA matching};

			\node (dec_match) [decision, below of=pro_match, yshift=-0.5cm] {Known IDFA?};

			\node (proatt_paid) [io_attribution, below of=dec_match, yshift=-0.5cm] {Attribute to paid campaign};

			\node (proatt_organic) [io_attribution, right of=dec_match, xshift=2cm] {Attribute to organic};

			\draw [arrow] (io_imp) -- (pro_click);
			\draw [arrow] (pro_click) -- (pro_clickatt);
			\draw [arrowdots] (pro_click) -- node[anchor=east] {} (pro_openapp);
			\draw [arrow] (pro_openapp) -- (pro_openappatt);
			\draw [arrowdots] (pro_organic.north) -- (pro_openapp.south);

			\draw [arrow] (pro_clickatt) -| (pro_match);
			\draw [arrow] (pro_openappatt) -- (pro_match);
			\draw [arrow] (pro_match) -- (dec_match);

			\draw [arrow] (dec_match.south) -- (proatt_paid.north);
			\draw [arrow] (dec_match) -- (proatt_organic);

		\end{tikzpicture}
	}
	\caption{How last-click attribution works when IDFA is available.
		The mobile measurement partner (MMP) uses the Identifier for Advertisers (IDFA) to identify the users that engaged with the mobile advertising campaign,
		and to separate those that installed the app organically (e.g., searching on the AppStore).
		This process will remain the same for the users that allow apps A and B to track them.
	}
	\label{fig:attribution_before}
\end{figure*}

\begin{figure*}[b!]
	\centering
	\resizebox{.61\textwidth}{!}{
		\begin{tikzpicture}[node distance=5cm, inner sep=7, outer sep=2]
			\huge
			\node (io_imp) [io_impression] {App A showing an ad promoting app B via campaign 1};
			\node (pro_click) [process, below of=io_imp, yshift=-1cm] {The user clicks on the ad, visits the App Store, and installs app B};
			\node (pro_openapp) [process, below of=pro_click, yshift=-0.5cm] {User opens app B for the first time};

			\node (pro_organic) [process_organic, below of=pro_openapp, yshift=-0.5cm] {User finds app B organically and installs it};

			\node (pro_openappatt1) [process, right of=pro_openapp, xshift=2cm] {App registers for AdNetwork Attribution};
			\node (pro_openappatt2) [process, left of=pro_openappatt1, xshift=12cm] {Update conversion value $v_i$};
			\node (dec_match) [decision, below of=pro_openappatt2, yshift=-0.5cm] {24 hours \\ passed?};
			\node (proatt) [io_attribution, below of=dec_match, yshift=-0.5cm] {Conversion value postback};

			\draw [arrow] (io_imp) -- (pro_click);
			\draw [arrowdots] (pro_click) -- node[anchor=east] {} (pro_openapp);
			\draw [arrowdots] (pro_organic) -- node[anchor=east] {} (pro_openapp);
			\draw [arrow] (pro_openapp) -- (pro_openappatt1);
			\draw [arrow] (pro_openappatt1) -- (pro_openappatt2);
			\draw [arrow] (pro_openappatt2) -- (dec_match);
			\draw [arrow] (dec_match.east) -| (20, -11.5) node[anchor=west, xshift=0.0cm, yshift=-2cm] {higher $v_i$} -- (pro_openappatt2.east);
			\draw [arrow] (dec_match) -- node[anchor=east, xshift=11cm] {Yes, wait randomly up to 24 hours}  (proatt);
		\end{tikzpicture}
	}
	\caption{Attribution when a user does not allow tracking via the App Tracking Transparency framwork.
		If the user does not allow tracking, then the attribution is done via conversion value.
		The conversion value postback contains various attributes about the network and campaign,
		but it does not have an ID for mapping the user and the advertising campaign that brought them.
		Application developers can track the conversion value for each user but not their origin.}
	\label{fig:attribution_after}
\end{figure*}
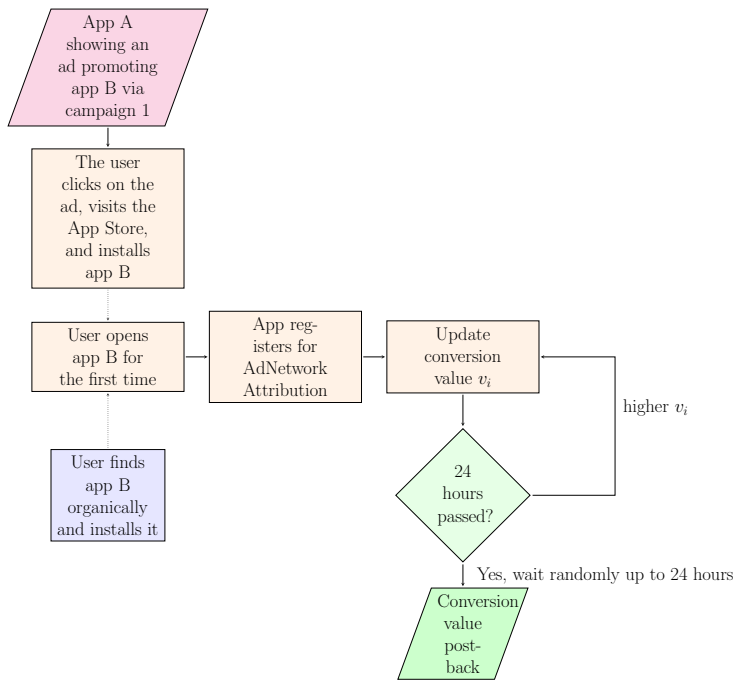

As a consequence, technological giants such as Google and Microsoft started to utilize privacy-preserving techniques~\cite{erlingsson2014rappor,ding2017collecting}.
Apple has previously introduced various privacy features~\cite{apple2017dp}, and recently they introduced their new version of the ad network API (SKAdNetwork 2.0) with support for a new framework called App Tracking Transparency (ATT).
Starting iOS 14.5, app developers cannot share any tracking identifier to advertising networks unless users allow it.
The App Tracking Transparency framework allows showing a pop-up dialog asking the user if they want to allow the application to track or not.
This privacy innovation has a profound impact on how ad campaigns' performance is measured. Without explicit consent from the user, it will not be possible to do last-click attribution.
Inevitably, the effectiveness of mobile advertising is affected, as the lack of the identifier affects how performance is measured and what types of ad personalization are available.
Figure~\ref{fig:attribution_before} presents how attribution works when IDFA is available, and
Figure~\ref{fig:attribution_after} presents what happens if a user does not allow tracking via the App Tracking Transparency framework, i.e., what happens when users allow tracking and when they do not.
The application developers have a  24 hour timer to report a new conversion value for the user.
When a new conversion value is set, the 24 hour timer is reset, allowing observing the player for another 24 hours.
Increasing the observation period gives more time to observe the player's actions, making conversion value schemas that rely on purchases feasible, as we will show later.

Let us introduce a hypothetical example. Suppose that an application developer promotes their app by showing a video and that a user gets interested in the ad, clicks on it, and installs the app.
The first time when the user opens the app, they can either allow tracking or not.
\begin{ex}
	If the user allows tracking, an identifier is shared to attribute the install to the clicked ad; and if they spend money on the application, the app developers can attribute the revenue to the ad campaign.
\end{ex}
\begin{ex}
	If the user does not allow tracking, the required identifier for attribution is not shared, but instead the advertising networks receive a postback with the campaign ID and a conversion value when it meets certain conditions.
\end{ex}


At a glance, the conversion values are a privacy-preserving mechanism proposed by Apple to measure an advertising campaign's performance without disclosing the user's origin.
In its core, conversion value separates the users into buckets.
The application developers are free to determine the bucket for each user based on the available information about them, e.g., they could use revenue, in-app events, retention, device type, and so on.
On the other hand, application developers do not know what campaign brought the user or if the user came organically, so this information cannot be incorporated into the conversion value.
Instead, Apple reports the count of users with the same conversion values per campaign via postbacks.
Essentially --- based on the currently available information --- the conversion values provide an ad-hoc privacy protection in the``hide-in-the-crowd' sense; such as
$k$-anonymity~\cite{sweeney2002k},
$l$-diversity~\cite{machanavajjhala2007diversity},
$t$-closeness~\cite{li2007t}
(instead of differential privacy~\cite{desfontaines2020sok} which has a formal privacy guarantee).
Apple released iOS 14.5 in April 2021, and its impact is already showing on advertising networks.
In the 2021 Q3 Snapchat earnings call, their CEO Evan Spiegel explained that one of the reasons why they missed their revenue guidance was the following:
``Our advertising business was disrupted by changes to iOS ad tracking that were broadly rolled out by Apple in June and July.
While we anticipated some degree of business disruption,
the new Apple provided measurement solution did not scale as we had expected,
making it more difficult for our advertising partners to measure and manage their ad campaigns for iOS ''\cite{snap_incq3}.
Snapchat stock plummeted more than 25\% after missing the earnings guidance.
Facebook comments on the impact of iOS during their 2021 Q3 earnings call:
``Overall, if it wasn’t for Apple’s iOS14 changes, we would have seen positive quarter-over-quarter
revenue growth. And while we and our advertisers will continue to feel the effect of these changes in
future quarters, we will continue working hard to mitigate them.''\cite{facebook_incq3}.
Twitter mentioned that ``It's still too early for Twitter to assess the long-term impact of Apple's privacy-related iOS changes, but the Q3 revenue impact was lower than expected, and we've incorporated an ongoing modest impact into our Q4 guidance''\cite{twitter_incq3}.
Finally, Alphabet commented that the impact of the iOS changes was ``a modest impact on YouTube revenues'' \cite{google_incq3}.
There is no doubt that marketing teams need a way to measure the performance of marketing campaigns under the new privacy-preserving mechanisms that Apple enforced starting iOS 14.5.
\paragraph{Contribution. }
This paper investigates various conversion value schemas in combination with revenue attribution functions.
Our contributions shed light on using the conversion values for attributing the revenue to the advertising campaigns. More specifically, our contributions are the following:
{\it i)} formalizing the problem of revenue attribution based on conversion values,
{\it ii)} finding the revenue attribution function which minimizes the attribution error for any conversion value schema,
{\it iii)} showing the revenue attribution quality of different conversion value schemas via back-testing on historical data.
\paragraph{Organization. }
In Section~\ref{sec:pre_rw} we provide the preliminaries for the conversion value schema \& revenue attribution and briefly review the relevant related work.
In Section~\ref{sec:mod} we present the models used for attributing the revenue to campaigns, provide a corresponding privacy analysis, and find theoretically the optimal revenue attribution function.
In Section~\ref{sec:exp} we show our experimental results and in Section~\ref{sec:conclusions} we conclude the paper.
\section{Background \& Related Work}\label{sec:pre_rw}
This section introduces the concepts and methods used in the rest of the paper, such as conversion value, Identifier for Advertisers (IDFA), last-click attribution, and user origin.
Moreover, we survey the related literature concerning both privacy and revenue attribution using conversion values.
For a general view on the challenges of privacy-centric digital advertising refer to G. Johnson et. al. \cite{johnson2021privacy}.
\subsection{Revenue Attribution}
Knowing the advertising campaign that brought a player to the game helps assign the revenue generated from the player to the campaign,
which is needed to measure ROI. App developers promote their apps in various ad networks.
Hence, a user might see ads for the same app in more than one network.
This leads to the question {\it what was the ad that caused the install?}
There are several approaches to answering this question~\cite{li2016attribution}.
For instance a top-down based approach is marketing mix modeling~\cite{chen2017_marketingmix}.
Other approaches are click-based methods such as first-click attribution~\cite{li2016attribution}, equal attribution~\cite{kannan2016path},
and last-click attribution~\cite{li2016attribution}.
There is an orthogonal research direction on how to improve click-based attribution methods.
In this paper, we focus on the last-click attribution since it is the most commonly used method in online advertising~\cite{dalessandro2012causally}.
Recent research suggests that in the mobile gaming industry only 9.5\% of the observed installs had impressions from more than one channel during a seven-day attribution window~\cite{syrjanen2019multi}.
In last-click attribution, we attribute the install to the last ad the user engaged with before/ installation.
Handling the last-click attribution is a complex problem, and often app developers delegate the task to attribution partners responsible for determining the last ad a user clicked.
\paragraph{User Origin. }
The origin of a user can be organic, paid advertisement, and cross-promotion.
The organic users have found the app in the App Store without previously engaging with an ad, for example, by searching for the app or scrolling through a list of popular apps, non-sponsored recommendations (e.g., App Store Best of 2021), and non-sponsored keywords.
Paid advertising includes, for example, social networks, search engines, or in-app ads.
Cross-promotion happens when users install the game by engaging with an ad from the same developer shown in their apps.
The ability to separate a user's origin allows companies to use the revenue from paid origin users to measure the ROI.
The separation of these origins allows modeling the app's virality (i.e., organic users invited by users from the paid origin) to include a share of the organics' revenue into the ROI calculation.
This paper focuses on understanding how conversion values help attribute users' revenue from the paid and organic origin.
\paragraph{Identifier for Advertisers. }
An identifier is required to know if a user came from a paid origin or organically.
In the Apple ecosystem, the identifier is called Identifier for Advertisers (IDFA), and its purpose is to allow tracking without disclosing the user's identity.
Starting iOS 14.5, users can set their preference for app tracking globally or per app.
The users may disable allowing apps to request to track system-wide, which means that the pop-up to give tracking consent is not shown.
If users allow apps to request to track, they can individually allow an application to know their IDFA.
The IDFA will serve its purpose only for the apps that the user gives system consent.
For last-click attribution to work, the user must give tracking consent in the app where the ad is shown and in the app that is being promoted.
\paragraph{Conversion Value. }
The conversion values are based on Apple's developer documentation of the \texttt{SKAdNetwork 2.0}~\cite{apple_doc_skadnetwork} and \texttt{update}-\texttt{ConversionValue}~\cite{apple_doc_cv}.
The conversion value is an integer $v\in [0, 63]$ that developers can set.
The conversion value is assigned for the first time when a user opens the app (i.e., not when the user installs the app).
Developers can increment the value within 24 hours of the last update. If there has been no update within 24 hours, the advertiser receives a postback of the install after a random time between 0 to 24 hours.
Even though theoretically this would allow for a two-month time window to update the conversion value, receiving the ad campaign's conversion value after two months is not very useful to guide advertising spend.
Practically, a period of up to seven days seems to be a maximum delay that makes sense, with many ad networks recommending much shorter windows (e.g., a 24 hour period)~\cite{conf_adj, conf_sing}.
The conversion value ranges from 0 to 63, and it is often modeled using a binary representation of six bits, where each bit may represent an action of the users as a logical condition (e.g., user passed tutorial, reached a certain level).
\begin{ex}
	A conversion value schema can use different bits for different things.
	An example of such schema would be using 2 bits to capture the days since the first
	opening (i.e., \texttt{00} initially, then \texttt{01}, \texttt{10}, \texttt{11}
	respectively after 1, 2, and 3 days) and 4 bits for in-app events, such as unlocking
	various app features and/or making purchases.
\end{ex}
\paragraph{Revenue Attribution using Conversion Values. }
The task of attributing revenue using conversion values is very recent.
To the best of our knowledge, our work is the first that formally investigates the task.
However, related work can be found on the Web.
An overview of the changes in \texttt{SKAdNetwork 2.0} is presented in~\cite{skadnetwork_101,apple_killed_idfa}.
Closer to our work is~\cite{algolift_prob_rev}, where the authors present two approaches that rely on the user's conversion value empirical conditional probability of belonging to a campaign.
The first approach is `winner takes all', which assigned the user (and its revenue) to the campaign with the highest empirical conditional probability.
The second approach is `probabilistic attribution',
which multiplies the user's revenue by the empirical conditional probability of coming from each campaign given the conversion value,
and sums it at the campaign level.
We use an equivalent approach in Equation~\ref{eq:sol}, but instead of attributing revenue using the user-campaign probabilities,
we use the expected revenue per conversion value multiplied by the count of conversion values per campaign and show that this method is optimal.
\subsection{Related Privacy Literature}
Many privacy preserving techniques were introduced in the last quarter century.
One of the most famous is $k$-anonimity~\cite{sweeney2002k} which requires that any user contained in the dataset cannot be distinguished from at least $k-1$ other users.
This was later improved by $l$-diversity~\cite{machanavajjhala2007diversity},
$t$-closeness~\cite{li2007t}, and $n$-confusion~\cite{stokes2012n}.
The main drawback of the methods described above is that they define anonymity as a property of the dataset.
Another wide-spread privacy mechanism is Differential Privacy~\cite{dwork2006differential} where anonymity is defined as a property of the process, making it resilient to any privacy attack based on background knowledge.
It was adapted to numerous scenarios, each requiring its own fine-tuning of the definition~\cite{desfontaines2020sok}.
These approaches are widely utilized in the industry as well as by organizations like
Google~\cite{erlingsson2014rappor}, Microsoft~\cite{ding2017collecting},
LinkedIn~\cite{kenthapadi2018pripearl}, Uber~\cite{chorus2018johnson},
and Apple~\cite{apple2017dp}.
We are not aware of any official well-detailed documentation concerning Apple's conversion value, only press releases, and blog posts~\cite{skadnetwork_101,apple_killed_idfa,algolift_prob_rev}.
Apple is not willing to reveal details about the mechanisms governing the conversion values. When Apple announced using differential privacy, they did it without telling crucial elements~\cite{tang2017privacy}.
Although this can also be seen as additional level of protection, it is well-known and widely believed that security and privacy by obscurity is never a good idea.
The reason originates from cryptography, where it is always assumed that the enemy knows the system being used~\cite{shannon1949communication}.
\section{Formalizing \& Analyzing Conversion Values}\label{sec:mod}
This section illustrates the problem, formalizes it, and analyzes it.
Our goal is to capture the scenario with all its details via a flexible mathematical model (e.g., it is adaptable for future changes concerning the conversion value schema),
as it is not certain how the conversion value schema is enforced.
Yet, based on our empirical observations (concerning the conversion value schema) and despite the intricate nature of the problem, our analysis lands itself on a simple solution concerning the optimal revenue attribution function,
which minimizes the difference between attributing revenue using conversion values and last-click attribution with IDFA.
The variables used in the paper are introduced individually in this section as well as summarized in Table~\ref{tab:params}.
\begin{table*}[t!]
	\begin{tabular}{p{.05\linewidth}|p{.8\linewidth}}
		Sym.                    & Meaning                                                                                                                                         \\
		\midrule
		$i$                     & User ID, in-between $1$ and $|U^d|$.                                                                                                            \\
		$d_i$                   & User's registration date: the first time a user opens the app.                                                                                  \\
		$t$                     & Number of days for the revenue to be accumulated (e.g., 3, 7, 14, 30, 90, etc.).                                                                \\
		$d$                     & The date when the conversion values are reported (sufficiently later than any $d_i$).                                                           \\
		$\alpha_i$              & User combined network and campaign ID: $\alpha=100\cdot n + c$. Note that $0\le c\le99$.                                                        \\
		$\beta$                 & The upper limit on $\alpha$ (i.e., $100\cdot n + c<\beta$). $\alpha=\beta$ corresponds to the organic users.                                    \\
		$r_i^t$                 & Accumulated revenue of the corresponding user $i$ for the first $t$ days after $d_i$.                                                           \\
		$\mathcal{U}_i$         & User features dataset (i.e., remaining information about the user).                                                                             \\
		$u_i^d$                 & $=(d_i, r_i^t, \alpha_i, \mathcal{U}_i)$, user data at $d$ where campaign IDs are known.                                                        \\
		$v_i^d$                 & Conversion value of user $i$ at $d$. Without subscript we mark the different conversion values.                                                 \\
		$\tilde{u}_i^d$         & $=(d_i, r_i^t, v_i^d, \mathcal{U}_i)$, user data when only conversion values are available instead of $\alpha_i$.                               \\
		$f(\cdot)$              & Conversion value schema or conversion value model (e.g., $f(u_i^d\setminus\{\alpha_i\})=v_i^d$).                                                \\
		$x_{v,\alpha}^d$        & $\in X^d$, the count of users in $v$ bucket at $d$ corresponding to $\alpha$.                                                                   \\
		$y_\alpha^t$            & Accumulated last-click attribution revenue for $\alpha$ based on the first $t$ days of the users.                                               \\
		$\tilde{U}_v^d$         & Set of all users (i.e., independently of $d_i$) with conversion value $v^d$, i.e., $\forall\tilde{u}_i\in\tilde{U}_v^d:v_i=v$.                  \\
		$\bar{r}_v^t$           & The average first $t$ days revenue of users in $\tilde{U}_{v}^d$ at $d$, i.e., $\bar{r}_v^t=\frac{\sum_{\tilde{U}_v^d}r_i^t}{|\tilde{U}_v^d|}$. \\
		$pr_p(\cdot)$           & Privacy preserving method with privacy threshold $p$.                                                                                           \\
		$\hat{x}_{v, \alpha}^d$ & $\in \hat{X}^d=pr_p(X^d)$, the conversion value counts after applying the privacy protection.                                                   \\
		$g_\alpha(\cdot)$       & Function to attribute the revenue of $\alpha$ at $d$. Input: $(\tilde{U}_v^d, \hat{x}_{v,\alpha})$.                                             \\
	\end{tabular}
	\caption{Summary of the variables used in the paper. }
	\label{tab:params}
\end{table*}
\subsection{Problem Illustration}
Initially, the app developers could distinguish between paid and organic users.
Thanks to IDFA and attribution methods like last-click attribution they could map the users to their origin, network, and advertising campaign.
A common practice for measuring the ROI is to group users in cohorts based on their registration date, origin, network, campaign, and country.
At the cohort level, one can aggregate the cost of acquiring the users and the revenue generated by them, which helps monitor ROI.
For simplicity, we note user IDs as $i\in\{1,2,\dots\}$ and the registration date of a user $i$ as $d_i$.
We are interested in $r_i^t$ which is the accumulated revenue of user $i$ from their registration $d$ until time $t$, so we must restrict ourselves to users with $d_i\le d-t$.
This is necessary; otherwise, the revenue attribution would become a prediction problem because we would not know $r_i^t$.%
\footnote{This is an important and exciting research question by itself and studied extensively~\cite{schmittlein1987counting,fader2005counting}.
	On the other hand, the problem studied in this paper (i.e., revenue attribution based on conversion values) attributes actual data from the user rather than forecasted data.}
The users satisfying this condition are captured as $u_i^d\in U_t^d$.
For convenience, we define the combination of network ID $n\in\mathbb{N}$ and ad campaign ID $c\in[0,99]$ to be $\alpha=100\cdot n+c$ because Apple restricts the amount of campaigns per network to 100.
We denote with $\beta$ the total number of different network and campaign combinations, hence $0\le\alpha<\beta$.
Note that the organic users correspond neither to any networks nor campaigns, hence, we capture them by setting their combined network and campaign ID to $\beta$.
Table~\ref{tab:original} presents the initial dataset when IDFA is available.
User-wise data is shown in Table~\ref{tab:userwise_PRE}, and Table~\ref{tab:campaignwise_PRE} shows the cumulative revenue $y_\alpha^t$ of the first $t$ days for each ad network \& campaign, which simplifies calculating the ROI.
Formally, this represents the data corresponding to $u_i^d$ as a tuple $\{d_i, r_i^t, \alpha_i, \mathcal{U}_i\}$.
The tuple includes the registration date,
the first $t$ day revenue generated by the user,
the user's origin, and --- for the sake of completeness --- it also contains $\mathcal{U}_i$ which captures any other related information about user $i$ such as event-level data within the app.

\begin{table}[t!]
	\begin{subtable}{\linewidth}\centering
		\begin{tabular}{c|c|cc|c}
			\toprule
			User ID $i$ & Revenue $r^t$ & Net. ID $n$ & Cam. ID $c$ & $\alpha_i$ \\
			\midrule
			$1$         & $0$ USD       & $4$         & $05$        & $405$      \\
			$2$         & $2.99$ USD    & $-$         & $-$         & $\beta$    \\
			$3$         & $0$ USD       & $3$         & $89$        & $389$      \\
			$\vdots$    & $\vdots$      & $\vdots$    & $\vdots$    & $\vdots$   \\
			$|U|$       & $4.99$ USD    & $1$         & $71$        & $171$      \\
			\bottomrule
		\end{tabular}
		\caption{Available user-wise data.}\label{tab:userwise_PRE}
	\end{subtable}%

	\begin{subtable}{\linewidth}\centering
		\begin{tabular}{c|cc|c}
			\toprule
			$\alpha$ & Net. ID $n$ & Cam. ID $c$ & Revenue $y^t$ \\
			\midrule
			$000$    & $0$         & $00$        & $245$ USD     \\
			$001$    & $0$         & $01$        & $92$ USD      \\
			$\vdots$ & $\vdots$    & $\vdots$    & $\vdots$      \\
			$099$    & $0$         & $99$        & $811$ USD     \\
			\midrule
			$100$    & $1$         & $00$        & $373$ USD     \\
			$\vdots$ & $\vdots$    & $\vdots$    & $\vdots$      \\
			$199$    & $1$         & $99$        & $373$ USD     \\
			\midrule
			$\vdots$ & $\vdots$    & $\vdots$    & $\vdots$      \\
			\midrule
			$\beta$  & $-$         & $-$         & $1639$ USD    \\
			\bottomrule
		\end{tabular}
		\caption{Available campaign-wise data.}\label{tab:campaignwise_PRE}
	\end{subtable}%

	\caption{Illustration of the user data which are available to add developers before Apple's App Tracking Transparency came out.}
	\label{tab:original}
\end{table}

\begin{table}[t!]
	\begin{subtable}{\linewidth}\centering
		\begin{tabular}{c|cccc}
			\toprule
			User ID $i$ & $1$     & $2$        & $\cdots$ & $|U|$      \\
			\midrule
			$r_i^t$     & $0$ USD & $2.99$ USD & $\cdots$ & $4.99$ USD \\
			$v_i$       & 0       & 6          & $\cdots$ & 63         \\
			\bottomrule
		\end{tabular}
		\caption{Available user-wise data.}\label{tab:userwise_POST}
	\end{subtable}%

	\begin{subtable}{\linewidth}\centering
		\begin{tabular}{c|cccc|ccc|c}
			\toprule
			$\alpha$ $\rightarrow$ & $000$    & $001$    & $\cdots$ & $099$    & $100$    & $\cdots$ & $199$    & $\cdots$ \\
			\midrule
			$v=0$                  & $19$     & $420$    & $\cdots$ & $88$     & $0$      & $\cdots$ & $36$     & $\cdots$ \\
			$v=1$                  & $355$    & $107$    & $\cdots$ & $31$     & $279$    & $\cdots$ & $151$    & $\cdots$ \\
			$v=2$                  & $329$    & $22$     & $\cdots$ & $34$     & $528$    & $\cdots$ & $2$      & $\cdots$ \\
			$\vdots$               & $\vdots$ & $\vdots$ & $\ddots$ & $\vdots$ & $\vdots$ & $\ddots$ & $\vdots$ & $\vdots$ \\
			$v=63$                 & $138$    & $346$    & $\cdots$ & $54$     & $7$      & $\cdots$ & $189$    & $\cdots$ \\
			\bottomrule
		\end{tabular}
		\caption{Available campaign-wise data.}\label{tab:campaignwise_POST}
	\end{subtable}%

	\caption{Illustration of the user data which are available to add developers after App Tracking Transparency came out. }
	\label{tab:new}
\end{table}

With the enforcement of App Tracking Transparency, the data presented in Table~\ref{tab:original} will not be available for the vast majority of the users, as explicit tracking system consent must be given.
Instead, the application developers have two tables available.
The first contains the conversion values $v_i$ and the revenues $r_i^t$ for all the app users, as presented in Table~\ref{tab:userwise_POST}.
The second contains the aggregate count of conversion values $X^d$ at time $d$, as shown in Table \ref{tab:campaignwise_POST}.%
\footnote{App developers do not receive the conversion values for $\beta$ (i.e., organic users) but can estimate them by subtracting the number of reported conversion values to the number of installs.}
Formally, when IDFA is not available, the user's tuple $u_i^d$ contains the same data, but instead of $\alpha_i$ the user $i$'s conversion value $v_i$ will be included.
We encapsulate this with $\tilde{u}_i^d=\{d_i, r_i, v_i, \mathcal{U}_i\}$.
The conversion value itself is computed from available user data via a conversion value schema $f$, i.e., $f(u_i^d\setminus\{\alpha_i\})=v_i$.\footnote{Note that our mathematical model slightly simplify the scenario visualized in Figure \ref{fig:attribution_after}, as we do neither consider the 24h delay from last conversion value update nor the additional 0-24h random delays, i.e., the conversion value count $X^d$ includes all users with $d_i\le d-t$. In contrast, our empirical analysis in Section \ref{sec:exp} does consider these, and follows further empirical observations obtained by interacting with the SKAdNetwork: any user belonging to $X^d$ are disregarded for any future $X^{d^\prime}$ where $d<d^\prime$.}
To ease the presentation of this paper we define three additional variables:
$\tilde{U}_v^d$ is the set of users with the same conversion value on day $d$,
$\bar{r}_v^t$ is the average revenue from users in $\tilde{U}_v^d$ within the first $t$ days,
and $g$ is the function to attribute the revenues based on the conversion value counts (i.e., $g$ approximates $y_\alpha^t$). In the rest of the paper, we slightly abuse the notations by leaving out superscript $dwhen it $ does not play a significant role.
\subsection{Privacy Protection}
The count of conversion values provides privacy protection in the form of `hide-in-the-crowd', as the campaign information does not contain user identifiers.
Individual users could still be connected with specific networks and campaigns if the size of a conversion value buckets are low.
For instance if only user $i$ has a specific conversion value then Table \ref{tab:new} would indeed reveal user $i$'s origin $\alpha_i$.
To overcome this problem, Apple proposed the privacy threshold $p$, a predefined (and currently unknown) value that provides further protection.%
\footnote{
	Apple's SKAdNetwork documentation~\cite{apple_doc_skadnetwork} mentions that \emph{``The postback may include a conversion value and the source app’s ID if Apple determines that providing the values meets Apple’s privacy threshold''}.}
On the other hand the documentation does not mention on which level the $p$ is enforced, e.g., only conversion value level (i.e., $|\tilde{U}_v^d|\ge p$), conversion value and campaign level (i.e., $x_{v,\alpha}^d\ge p$), country level, etc.).
In practice, Apple will not report the count of users in the conversion values where there are less than $p$ users, and instead, those counts will be reported as \texttt{null}.
Moreover, the users with such a conversion value are not discarded.
Instead, the set of conversion values is extended with \texttt{null}, i.e., $v\in\{\texttt{null}, 0, 1, \dots, 63\}$ which aggregates all the users from below the threshold conversion values.
\begin{align}
	\label{eq:privacy}
	pr_p(X)=\hat{X}=
	\begin{cases}
		\hat{x}_{v,\alpha}=
		\begin{cases}
			x_{v,\alpha}  & \text{if } \sum_U\mathbbm{1}(v_i=v) \ge p \\
			\texttt{null} & \text{otherwise}
		\end{cases} \\
		\hat{x}_{\texttt{null},\alpha}=\sum_v\mathbbm{1}(\hat{x}_{v,\alpha}=\texttt{null})\cdot x_{v,\alpha}
	\end{cases}
\end{align}
Formally, our interpretation of the privacy mechanism (based on what we experience by interacting with the corresponding ecosystem post App Tracking Transparency) is defined in Equation \ref{eq:privacy}, where $\mathbbm{1}$ is the indicator function.
This mechanism is similar to $k$-anonymity~\cite{sweeney2002k}, which requires all users to be indistinguishable from at least $k-1$ other users.
On the other hand, it does not satisfy that because the condition is not enforced on $\hat{x}_{\texttt{null},\alpha}$.
\begin{ex}
	For instance, if there is a single user (e.g., $i$) with a particular conversion value (e.g., $v_i$),
	for all $\alpha$ the values $\hat{x}_{v_i,\alpha}$ are set to \texttt{null}.
	Moreover, if we assume that the size of all other conversion values are above $p$,
	then $\hat{x}_{\texttt{null},\alpha}=0$ for all $\alpha$ except for $\alpha_i$ in which case it is 1.
	Consequently, user $i$ is not similar to $p-1$ other users as they can be singled out.
\end{ex}
\subsection{Revenue Attribution Functions}
\label{sec:th}
The revenue attribution function $g$ plays a central role in our research, as we want to approximate the actual campaign-wise revenues $y_\alpha^t$ via the conversion values.
This attribution error minimization problem is shown in Equation~\ref{eq:problem}.
Although $f$ is not explicit in the formula to be minimized, it defines $\tilde{U}_{v}^d$ as it contains users with $f(u_i^d\setminus \{\alpha_i\})=v_i^d$.
\begin{equation}
	\label{eq:problem}
	\min_f\left[\sum_\alpha\left(\sum_vg\left(
		\tilde{U}_v^d,\hat{x}_{v,\alpha}^d\right)-y_\alpha^t\right)^2\right]
\end{equation}
First, instead of focusing on $f$, we show the optimal $g$ when there is no privacy threshold for conversion values (i.e., when $p<2$).
When $p=0$ it is meaningless, and when $p=1$ it only changes the 0 values to \texttt{null}, making no real difference between $X$ and $\hat{X}$.
\begin{theorem}
	\label{th:revenue}
	Only based on $\tilde{U}$ (i.e., without any prior background knowledge about the distribution of users corresponding to any $\alpha$)
	and if $p<2$ (i.e., we assume $\hat{x}_{v,\alpha}^d=x_{v,\alpha}^d$ so there is no additional privacy protection) then independently of $f$,
	the attribution function defined in Equation~\ref{eq:sol} minimizes Equation~\ref{eq:problem}.
	\begin{equation}
		\label{eq:sol}
		g_{\alpha}\left(\tilde{U}_v^d, x_{v,\alpha}^d\right)=x_{v,\alpha}^d\cdot\bar{r}_v^t
	\end{equation}
\end{theorem}
\begin{proof}
	See in the Appendix.
	\hfill\qedhere
\end{proof}
Now we relax our initial condition about $p$ and focus on the case when the privacy threshold is applied (i.e., when $p\ge2$).
The exact privacy preserving mechanism used by Apple is unknown, therefore we are using Equation \ref{eq:privacy} based on empirical available post-back data.
The revenue attribution function defined in Equation \ref{eq:sol} does not consider the \texttt{null} bucket.
To account for the \texttt{null} bucket, we propose two attribution functions in the form of Equation \ref{eq:rev_all},
where $\fbox{$\phantom{12}$}$ should be filled accordingly.
\begin{equation}
	\label{eq:rev_all}
	g_{\alpha}\left(\tilde{U}_v^d, x_{v,\alpha}\right)=
	\begin{cases}
		\bar{r}_v^t\cdot\hat{x}_{v,\alpha}                                 & \text{if } \hat{x}_{v,\alpha}\not=\texttt{null} \\
		\bar{r}_v^t\cdot\fbox{$\phantom{12}$}\cdot\sum_U\mathbbm{1}(v_i=v) & \text{otherwise}
	\end{cases}
\end{equation}
\paragraph{Uniform Revenue Attribution (\textbf{U}). }
Distributing the revenue uniformly across all possible networks and campaigns, i.e., $\frac1\beta$ should fill $\fbox{$\phantom{12}$}$ in Equation \ref{eq:rev_all}.
This function is used as a pessimistic baseline because it does not use any information from $\hat{X}^d$.
\paragraph{Null-based Revenue Attribution (\textbf{N}). }
Distributing the revenue based on the empirical distribution defined by the \texttt{null} bucket, i.e.,
$\frac{\hat{x}_{\texttt{null},\alpha}}	{\sum_\alpha\hat{x}_{\texttt{null},\alpha}}$
should fill $\fbox{$\phantom{12}$}$ in Equation \ref{eq:rev_all}.
This function is based on the `sum' of the distribution corresponding to conversion values below the threshold $p$.
Although we have no prior background information about the user distributions within the conversion values, we can still utilize \texttt{null} bucket for those below $p$.
\begin{theorem}
	\label{th:revenueP}
	Only based on $\tilde{U}$ (e.g., without any prior background knowledge about the distribution of users corresponding to any $\alpha$) for any $f$,
	the attribution function defined in Equation~\ref{eq:rev_all} minimizes Equation~\ref{eq:problem} where $\fbox{$\phantom{12}$}$ is a convex combination of \textbf{U} and \textbf{N}.
\end{theorem}
\begin{proof}
	See in the Appendix.
	\hfill\qedhere
\end{proof}
\section{Experiments}\label{sec:exp}
This section introduces several conversion value schemas and  shows the corresponding empirical results combined with the introduced revenue attribution functions.
\paragraph{Setup. }
We experiment using data from a globally launched free-to-play mobile game developed by one of the biggest mobile game developer's. 
To generate the ground truth dataset, we used six months of historical data from cohorts with revenue matured up to 90 days (i.e., players that have played at least 90 days).
The dataset includes more than 500K paid players which constitutes a significant share of players in the time window.
The users correspond to 213 campaigns scattered across 7 networks, hence $\beta=214$ (i.e., 213 plus organic).
Using historical data allows us to compare the attributed revenue with the actual data from last-click attribution.
We calculate the conversion value for each user in the dataset according to the schema we want to evaluate.
Because neither the exact privacy threshold nor the level (e.g., globally, country-wise, etc.) is known as of now,
we use Equation~\ref{eq:privacy} with different values of $p$ and with country-level privacy protection.
We build the matrix $\hat{X}^d$ eight times separately: six for countries with the largest user bases and two for the rest grouped randomly.
Instead of using the registration date of the players, we use weekly cohorts, meaning that the matrix $X^d$ contains the sum of counts of daily conversion values per week starting from Monday.
We calculate the error per week and then aggregate the error from different weeks using a weighted average where the weight is the week's revenue.
The experiments were implemented in Python and ran in a single machine with 64 vCPU and 512 GB of RAM.
\subsection{Conversion value schemas}\label{sec:cv_scheme}
\begin{figure}[t!]
	\centering
	\includegraphics[width=0.67\textwidth]{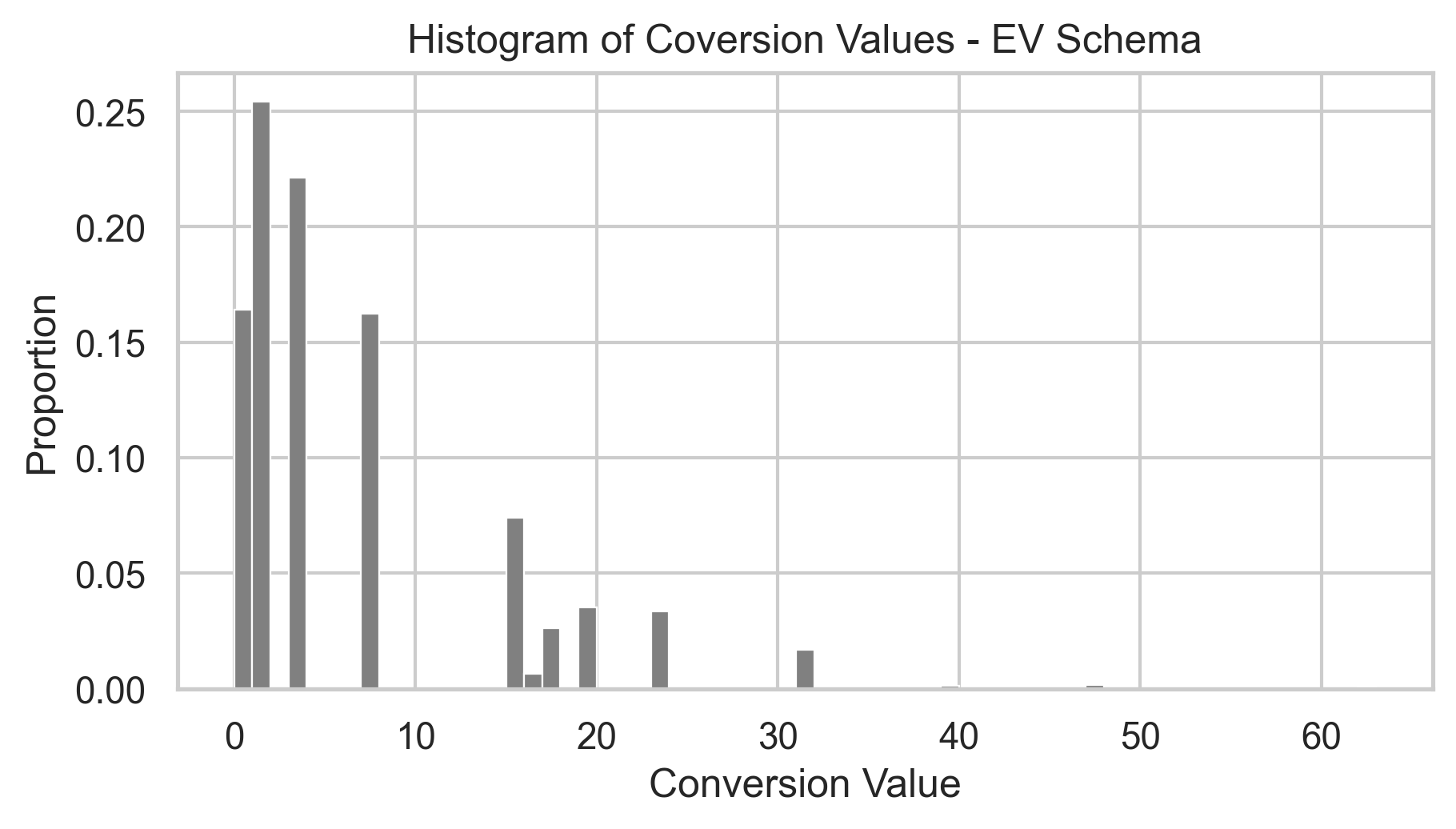}
	\caption{Histogram of conversion values using day 0 event-based (\textbf{EV}) schema when $p=0$. The plot is normalized so that the sum of bars is 1.}
	\label{fig:ev_schema}
\end{figure}

In Section \ref{sec:th} we described the revenue attribution function $g$ and showed theoretically which is the optimal.
Concerning conversion value schema $f$ we do not pursue this direction.
Rather, we assign meaning to each of the six bits of the conversion values
by defining three types of bits:
{\it T} bits used for time (i.e., day),
{\it V} bits used for revenue, and
	{\it C} bits used for a logical condition
(e.g., data captured within $\mathcal{U}$: device is tablet or smartphone, user passed tutorial, user reached a certain level, etc.).
Using these bits we specify various conversion value schemas.
It is worth mentioning that schemas using data beyond the registration day (i.e., day 0) are challenging in practice because they depend on the user coming back to play within 24 hours and to update the conversion value while the player is using the application.
Although this was not utilized in our theoretical results concerning the revenue attribution function $g$,
the experiments about the conversion value schema $f$ take into account the fact that the user needs to be active in a 24 hour time window since the last update and the new conversion value must be higher than previous one.
Next we define five conversion value schemas.
\paragraph{Day 0 event-based (\textbf{EV}). }
Using data from $\mathcal{U}_i$,
we encode six actions taken by the user during their first day of using the app (i.e., {\it CCCCCC}), each taken action corresponding to one bit (e.g., finishing the tutorial is bit $0$, reaching certain level is $1$, etc.).
Figure \ref{fig:ev_schema} shows an example of the frequency distribution of conversion values obtained using an \textbf{EV} schema\footnote{Some networks have a limitation of using just the first 24 hours of data to fix the conversion value for certain types of campaign optimizations (e.g., Facebook~\cite{link2}). For those networks, the only realistic schema would be the \textbf{EV}.}.
\paragraph{Rolling Revenue \& Rolling Purchase Count (\textbf{RR} \& \textbf{RI}). }
Both rolling schemas are utilizing some bits {\it T} for keeping track of the days that
have passed from the first opening. The remaining bits are defined by the purchases:
\textbf{RR} uses bits {\it V} for bucketing the actual revenue while \textbf{RI} uses bits {\it C} for bucketing the purchase counts of the user during the observation period, i.e., the first accumulates the total value of purchases while the latter counts how many purchases happened.
Users without revenue are assigned to the zero bucket, and those with revenue are distributed uniformly based on their revenue.
For example, \textbf{D7 RR} is defined as {\it TTTVVV}, where {\it T} bits capture day 0-7 and {\it V} bits are based on the current user's revenue.
In order to not disclose business confidential information, plots on revenue schemas could not be provided.
\paragraph{Uniform distribution (\textbf{UD}). }
Distributing users in conversion values at random.
This schema is used as a pessimistic baseline because it does not use any information of the user.
\paragraph{Perfect life time value (\textbf{PV}).}
Using six {\it V} bits to bucket users based on the future cumulative revenue of the user.
This is a hypothetical schema as it uses data which is not available in practice~\cite{malthouse2005can}.
For example, \textbf{D30 PV} is defined as {\it VVVVVV}, where bits are based on the user's cumulative revenue until day 30.
The schema serves as an optimistic baseline because it places users in a manner that their revenue is close to the conversion value's expected revenue.
\subsection{Results}
\begin{table*}[t]
	\begin{subtable}{\linewidth}\centering
		\begin{tabular}{c|c|cc|cc|cc}
			\toprule
			                        & $p=0$            & \multicolumn{2}{c|}{$p=2$} & \multicolumn{2}{c|}{$p=10$} & \multicolumn{2}{c}{$p=100$}                                        \\
			\midrule
			\backslashbox{$f$}{$g$} & Eq. \ref{eq:sol} & \textbf{U}                 & \textbf{N}                  & \textbf{U}                  & \textbf{N} & \textbf{U} & \textbf{N} \\
			\midrule
			\textbf{D30 PV}         & \boxed{0}        & \boxed{0}                  & 1                           & \boxed{0}                   & -2         & \boxed{0}  & 28         \\
			\midrule
			\textbf{EV}             & -42              & -25                        & -25                         & -2                          & -8         & 15         & 17         \\
			\textbf{D1 RR}          & -32              & -19                        & -19                         & -5                          & -5         & 18         & 24         \\
			\textbf{D1 RI}          & -34              & -20                        & -20                         & 0                           & -3         & 17         & 26         \\
			\textbf{D3 RR}          & -21              & -9                         & -8                          & -15                         & -10        & 11         & 35         \\
			\textbf{D3 RI}          & -25              & -12                        & -11                         & -7                          & -17        & 6          & 33         \\
			\textbf{D7 RR}          & -17              & -4                         & -4                          & -14                         & -6         & -7         & 25         \\
			\textbf{D7 RI}          & -21              & -8                         & -7                          & -7                          & -6         & -6         & 32         \\
			\midrule
			\textbf{UD}             & -62              & -43                        & -43                         & -15                         & -15        & 10         & 10         \\
			\bottomrule
		\end{tabular}
		\caption{Attribution benchmark for cumulative \emph{campaign revenue}.}
		\label{tab:schemes_at_g_campaign}
	\end{subtable}
	\begin{subtable}{\linewidth}\centering
		\begin{tabular}{c|c|cc|cc|cc}
			\toprule
			                        & $p=0$            & \multicolumn{2}{c|}{$p=2$} & \multicolumn{2}{c|}{$p=10$} & \multicolumn{2}{c}{$p=100$}                                        \\
			\midrule
			\backslashbox{$f$}{$g$} & Eq. \ref{eq:sol} & \textbf{U}                 & \textbf{N}                  & \textbf{U}                  & \textbf{N} & \textbf{U} & \textbf{N} \\
			\midrule
			\textbf{D30 PV}         & \boxed{0}        & \boxed{0}                  & 2                           & \boxed{0}                   & 7          & \boxed{0}  & 46         \\
			\midrule
			\textbf{EV}             & -32              & -15                        & -15                         & 7                           & 7          & 23         & 34         \\
			\textbf{D1 RR}          & -24              & -11                        & -10                         & 2                           & 5          & 28         & 40         \\
			\textbf{D1 RI}          & -26              & -12                        & -11                         & 8                           & 14         & 26         & 43         \\
			\textbf{D3 RR}          & -15              & -2                         & -1                          & -11                         & -1         & 15         & 50         \\
			\textbf{D3 RI}          & -17              & -5                         & -4                          & -2                          & 2          & 9          & 51         \\
			\textbf{D7 RR}          & -11              & 1                          & 2                           & -13                         & 1          & -6         & 44         \\
			\textbf{D7 RI}          & -13              & 0                          & 1                           & -5                          & 6          & -6         & 50         \\
			\midrule
			\textbf{UD}             & -50              & -31                        & -31                         & -3                          & -3         & 11         & 11         \\
			\bottomrule
		\end{tabular}
		\caption{Attribution benchmark for cumulative \emph{network revenue}.}
		\label{tab:schemes_at_g_network}
	\end{subtable}
	\caption{Attribution benchmark for 30 days of cumulative revenue.
		The error metric is normalized with \textbf{D30 PV} combined with \textbf{U} (also noted with a box).
		A negative value means worse than baseline and a positive value means better than baseline.
	}
	\label{tab:schemes_at_g}
\end{table*}
We want to know how the introduced conversion value schemas presented in Section~\ref{sec:cv_scheme} perform in revenue attribution.
We experimented by backtesting on past data, hence the ground truth is available via the data reported by our Mobile Measurement Partner.
This allows us to measure the revenue attribution error of various conversion value schemas.
Our results are presented in Table \ref{tab:schemes_at_g} where the prefix in the first column shows the number of bits used for time, e.g., D1 corresponds to 1 bit (one day).
As described in Section~\ref{sec:exp}, the revenue attribution was calculated on a weekly basis, so the results are compared using the average error for all the weeks.
From Theorem \ref{th:revenueP} we know that the optimal revenue attribution function is a convex combination of uniform \textbf{U} and \texttt{null}-based empirical \textbf{N}.
Because the exact combination is unknown, we consider both separately.
Table~\ref{tab:schemes_at_g} shows the errors for attributing the cumulative revenue for 30 days.
It shows that the best conversion value schemas use the observed users' revenues.
The attribution errors are normalized with the hypothetical best case \textbf{D30 PV} with \textbf{U} for every privacy parameter separately (also marked with a box).
For example, in Table~\ref{tab:schemes_at_g_campaign}, when $p=2$, the conversion value schema \textbf{D7 RR} combined with \textbf{N} is $4\%$ worse than the error of \textbf{D30 PV} with \textbf{U}.
As $p$ increases, the \textbf{EV} and \textbf{UD} schemas' performance gets closer to the rest of the schemas because a high privacy threshold applied to the revenue-based schema sets most of the spending user's conversion value to $null$.
As expected, the baseline schema \textbf{D30 PV} error is smaller than all other when there is no privacy threshold and \textbf{UD} performs the worst.
Looking at the results for \textbf{RR} and \textbf{RI} with low privacy threshold (i.e., $p\le2$) in Table~\ref{tab:schemes_at_g} we see that using a more extended period than the first 24 hours of gameplay reduces the attribution error.
Intuitively, it takes some time for players to try the game, and they will  start buying once they consider that it is worth it -- which rarely happens in the first day of gameplay.
\textbf{RR} and \textbf{RI} work well if there are enough players spending during the observed period, because it helps separating non-spenders from spenders, and then placing spenders in buckets based on their spending.
However, higher privacy thresholds affect the quality of this schema because observing the players for a few days cannot correctly separate players into different conversion values.
Most players will have a conversion value of zero, and those that do not will likely fall below the privacy threshold.
The results suggest that schemas that separate spenders and non-spenders and group users based on their spending are the most helpful for revenue attribution.
That is why {\it rolling} schemas that include bits for carrying the count of days perform much better than the \textbf{EV} or \textbf{UD} schema.
\begin{figure}[t!]
	\centering
	\includegraphics[width=0.67\textwidth]{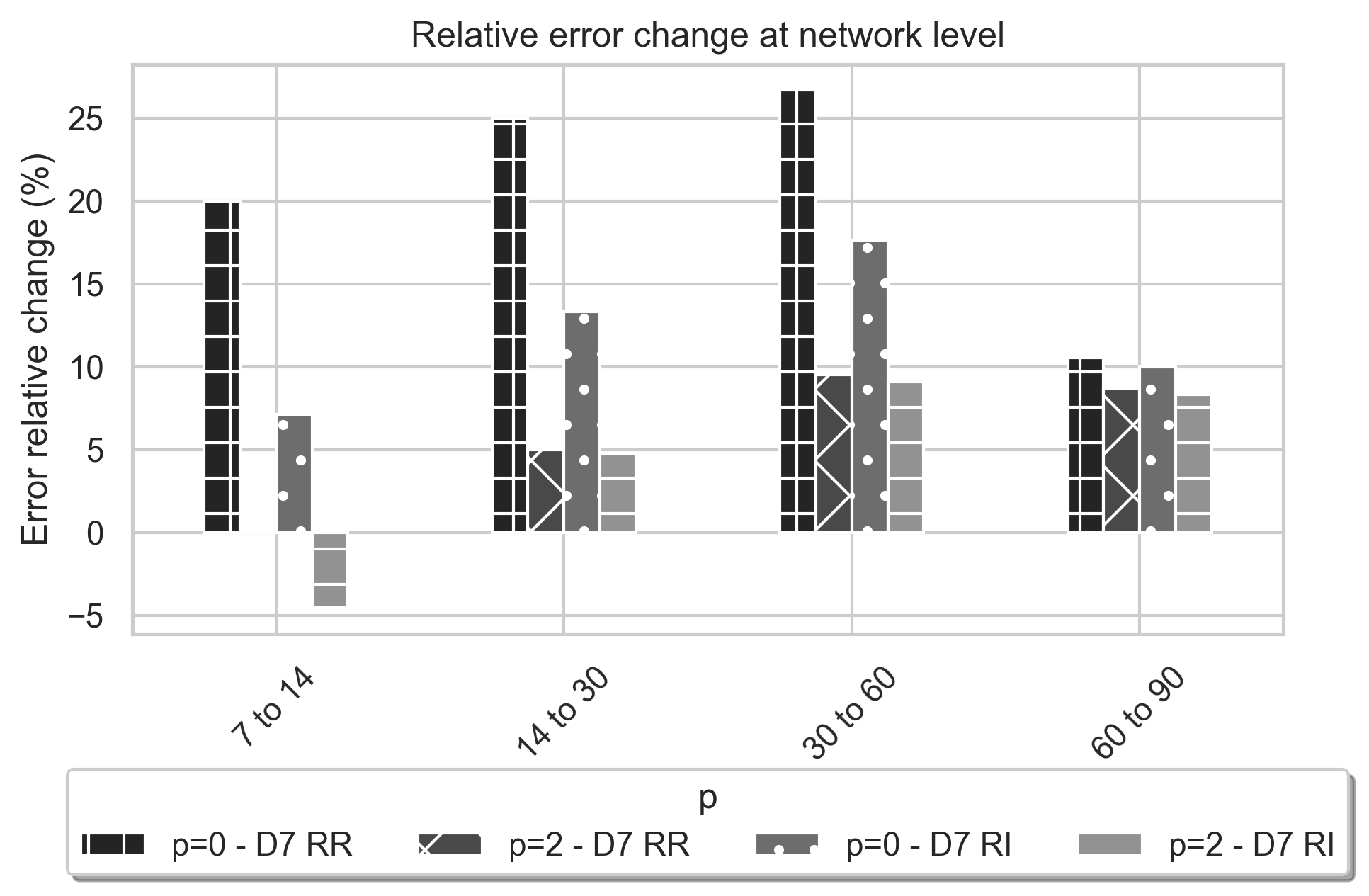}
	\caption{The attribution error grows as revenue matures.}
	\label{fig:att_error}
\end{figure}
Figure \ref{fig:att_error} visualizes the change in attribution error according to the revenue window. Windows of 7 to 14 days, 14 to 30 days, 30 to 60 days and 60 to 90 days are presented.
As we attribute cumulative revenue in more extended time, user behavior has more chances to diverge.
Using just the first days of data to group users does not mean that the users will have the same journey in the app.
Some users will keep playing without buying, and most players who buy will only do it once.
The users with a higher number of purchases will spend at a different pace, and their spending will vary from hundreds to thousands of dollars.
Finally, bucketing spenders based on early signals to group them based on long revenue windows of time is very challenging.
\section{Conclusion}
\label{sec:conclusions}
This paper focuses on using conversion values to attribute revenue to advertising campaigns. The conversion value schema is a novel privacy-preserving method for optimizing ad campaigns introduced by Apple in \texttt{SKAdNetwork 2.0} as part of the App Tracking Transparency framework, which was enforced in iOS 14.5 and later.
Instead of allowing advertisers to use IDFA by default, the user must give permission explicitly, impacting how the mobile measurement partner attributes revenue to advertising campaigns.
To the best of our knowledge, our work is the first to rigorously formalize and investigate the conversion values for revenue attribution.
We find the optimal revenue attribution function, and through various experiments we shed light on how different conversion value schemas perform in revenue attribution.
Based on empirical evaluation on real-world data we postulate that the best conversion value schema is the one that relies on revenue and is able to separate spenders and non-spenders.
\paragraph{Limitations. }
Our work barely scratches the surface of Apple's conversion value schema,
and we hope that it motivates the industry and academia to formally investigate the best ways to use conversion values.
We focused on using conversion values for attributing revenue in free-to-play games where more than 95\% of the users do not spend money.
It is likely that the conversion value schemas in other kinds of applications (e.g., subscription, ads driven)
and tasks (e.g., ads campaign optimization, real-time bidding) perform differently.
The major limitation of our work is that the rules of the privacy threshold are not clear.
If we would know the definition of the privacy threshold in detail we could simulate the conversion values better.
\paragraph{Future Work. }
We foresee many directions in which this line of research can continue. For instance:
\begin{itemize}
	\item The Effects of Opt-ins:
	      It is reasonable to assume that some users give permission to track in both the app that shows the ad as well as the app that gets installed~\cite{link1,krafft2017permission}.
	      In that case, the exact network and campaign information is available.
	      Hence, it might be possible to either estimate campaign revenue solely based on these users, or use the data from these users to improve the estimation that includes the opt-out users.
	\item Diagnosing revenue attribution:
	      Another promising research area is investigating how to measure the revenue attribution quality without knowing the ground truth.
	\item Optimal campaign structure:
	      Since the conversion values can be reported as \texttt{null} when the number of installs for a campaign has not reached the privacy threshold, it would be interesting to study what is the optimal number of campaigns to run to make the user acquisition operations as efficient as possible.
\end{itemize}
\bibliographystyle{plain}
\bibliography{conversion-value}

\begin{thebibliography}{10}

\bibitem{conf_adj}
Adjust.
\newblock Conversion value management, 2021.

\bibitem{algolift_prob_rev}
Algolift.
\newblock Probabilistic attribution for ios14 — a deeper dive, 2020.

\bibitem{google_incq3}
Alphabet.
\newblock Alphabet q3 2021 earnings call, 2021.

\bibitem{apple_doc_skadnetwork}
Apple.
\newblock Skadnetwork | apple developer documentation, 2020.

\bibitem{apple_doc_cv}
Apple.
\newblock updateconversionvalue(\_:) $|$ apple developer documentation, 2020.

\bibitem{blattberg1991interactive}
Robert~C Blattberg and John Deighton.
\newblock Interactive marketing: Exploiting the age of addressability.
\newblock {\em Sloan management review}, 33(1):5--15, 1991.

\bibitem{ccpa}
State California.
\newblock California consumer privacy act.
\newblock {\em OBLIGATIONS ARISING FROM PARTICULAR TRANSACTIONS}, 1798.100 -
  1798.199, 2018.

\bibitem{eurobarometer_eprivacy}
European Commission.
\newblock e-privacy.
\newblock {\em Flash Eurobarometer}, 443, 2016.

\bibitem{dalessandro2012causally}
Brian Dalessandro, Claudia Perlich, Ori Stitelman, and Foster Provost.
\newblock Causally motivated attribution for online advertising.
\newblock In {\em Proceedings of the sixth international workshop on data
  mining for online advertising and internet economy}, pages 1--9, 2012.

\bibitem{desfontaines2020sok}
Damien Desfontaines and Bal{\'a}zs Pej{\'o}.
\newblock Sok: Differential privacies.
\newblock {\em Proceedings on Privacy Enhancing Technologies},
  2020(2):288--313, 2020.

\bibitem{ding2017collecting}
Bolin Ding, Janardhan Kulkarni, and Sergey Yekhanin.
\newblock Collecting telemetry data privately.
\newblock In {\em Advances in Neural Information Processing Systems}, 2017.

\bibitem{dwork2006differential}
Cynthia Dwork.
\newblock Differential privacy.
\newblock In {\em Proceedings of the 33rd international conference on Automata,
  Languages and Programming}. ACM, 2006.

\bibitem{erlingsson2014rappor}
{\'U}lfar Erlingsson, Vasyl Pihur, and Aleksandra Korolova.
\newblock {RAPPOR}: Randomized aggregatable privacy-preserving ordinal
  response.
\newblock In {\em Proceedings of the 2014 ACM SIGSAC conference on computer and
  communications security}. ACM, 2014.

\bibitem{link2}
Facebook.
\newblock Preparing our partners for ios 14: Impacts to app advertisers and
  developers, 2021.

\bibitem{facebook_incq3}
Inc. Facebook.
\newblock Facebook, inc. (fb) third quarter 2021 results conference call, 2021.

\bibitem{fader2005counting}
Peter~S Fader, Bruce~GS Hardie, and Ka~Lok Lee.
\newblock “counting your customers” the easy way: An alternative to the
  pareto/nbd model.
\newblock {\em Marketing science}, 24(2):275--284, 2005.

\bibitem{link1}
Flurry.
\newblock ios 14.5 opt-in rate - daily updates since launch, 2021.

\bibitem{us_ntia}
Rafi Goldberg.
\newblock Most americans continue to have privacy and security concerns, ntia
  survey finds, 2018.

\bibitem{chen2017_marketingmix}
Google.
\newblock Challenges and opportunities in media mix modeling, 2017.

\bibitem{hughes2014multiple}
Gabriel Hughes and Damien Allison.
\newblock Multiple attribution models with return on ad spend, July~22 2014.
\newblock US Patent 8,788,339.

\bibitem{snap_incq3}
SNAP INC.
\newblock Snap inc. q3 2021 transcript, 2021.

\bibitem{johnson2021privacy}
Garrett Johnson, Julian Runge, and Eric Seufert.
\newblock Privacy-centric digital advertising: Implications for research.
\newblock {\em Available at SSRN 3947290}, 2021.

\bibitem{chorus2018johnson}
Noah Johnson, Joseph~P. Near, Joseph~M. Hellerstein, and Dawn Song.
\newblock Chorus: Differential privacy via query rewriting.
\newblock {\em arXiv preprint arXiv:1809.07750}, 2018.

\bibitem{kannan2016path}
PK~Kannan, Werner Reinartz, and Peter~C Verhoef.
\newblock The path to purchase and attribution modeling: Introduction to
  special section, 2016.

\bibitem{kenthapadi2018pripearl}
Krishnaram Kenthapadi and Thanh~TL Tran.
\newblock Pripearl: A framework for privacy-preserving analytics and reporting
  at linkedin.
\newblock In {\em Proceedings of the 27th ACM International Conference on
  Information and Knowledge Management}. ACM, 2018.

\bibitem{skadnetwork_101}
Yonatan Komornik.
\newblock Skadnetwork 101: What is it? what does it mean for you?, 2020.

\bibitem{krafft2017permission}
Manfred Krafft, Christine~M Arden, and Peter~C Verhoef.
\newblock Permission marketing and privacy concerns—why do customers (not)
  grant permissions?
\newblock {\em Journal of interactive marketing}, 39:39--54, 2017.

\bibitem{li2016attribution}
Hongshuang Li, PK~Kannan, Siva Viswanathan, and Abhishek Pani.
\newblock Attribution strategies and return on keyword investment in paid
  search advertising.
\newblock {\em Marketing Science}, 35(6):831--848, 2016.

\bibitem{li2007t}
Ninghui Li, Tiancheng Li, and Suresh Venkatasubramanian.
\newblock t-closeness: Privacy beyond k-anonymity and l-diversity.
\newblock In {\em 2007 IEEE 23rd International Conference on Data Engineering},
  pages 106--115. IEEE, 2007.

\bibitem{machanavajjhala2007diversity}
Ashwin Machanavajjhala, Daniel Kifer, Johannes Gehrke, and Muthuramakrishnan
  Venkitasubramaniam.
\newblock l-diversity: Privacy beyond k-anonymity.
\newblock {\em ACM Transactions on Knowledge Discovery from Data (TKDD)},
  1(1):3--es, 2007.

\bibitem{malthouse2005can}
Edward~C Malthouse and Robert~C Blattberg.
\newblock Can we predict customer lifetime value?
\newblock {\em Journal of interactive marketing}, 19(1):2--16, 2005.

\bibitem{regulation2016regulation}
Protection Regulation.
\newblock Regulation (eu) 2016/679 of the european parliament and of the
  council.
\newblock {\em REGULATION (EU)}, 679, 2016.

\bibitem{schmittlein1987counting}
David~C Schmittlein, Donald~G Morrison, and Richard Colombo.
\newblock Counting your customers: Who-are they and what will they do next?
\newblock {\em Management science}, 33(1):1--24, 1987.

\bibitem{apple_killed_idfa}
Eric~Benjamin Seufert.
\newblock Apple killed the idfa: A comprehensive guide to the future of mobile
  marketing, 2020.

\bibitem{shankar2009mobile}
Venkatesh Shankar and Sridhar Balasubramanian.
\newblock Mobile marketing: A synthesis and prognosis.
\newblock {\em Journal of interactive marketing}, 23(2):118--129, 2009.

\bibitem{shankar2016mobile}
Venkatesh Shankar, Mirella Kleijnen, Suresh Ramanathan, Ross Rizley, Steve
  Holland, and Shawn Morrissey.
\newblock Mobile shopper marketing: Key issues, current insights, and future
  research avenues.
\newblock {\em Journal of Interactive Marketing}, 34:37--48, 2016.

\bibitem{shannon1949communication}
Claude~E Shannon.
\newblock Communication theory of secrecy systems.
\newblock {\em The Bell system technical journal}, 28(4):656--715, 1949.

\bibitem{conf_sing}
Singular.
\newblock Skadnetwork model configuration faq, 2021.

\bibitem{stokes2012n}
Klara Stokes and Vicen{\c{c}} Torra.
\newblock n-confusion: a generalization of k-anonymity.
\newblock In {\em Proceedings of the 2012 Joint EDBT/ICDT Workshops}, pages
  211--215, 2012.

\bibitem{sweeney2002k}
Latanya Sweeney.
\newblock k-anonymity: A model for protecting privacy.
\newblock {\em International Journal of Uncertainty, Fuzziness and
  Knowledge-Based Systems}, 10(05):557--570, 2002.

\bibitem{syrjanen2019multi}
Joonas Syrj{\"a}nen et~al.
\newblock Multi-touch attribution in the mobile gaming industry.
\newblock 2019.

\bibitem{tang2017privacy}
Jun Tang, Aleksandra Korolova, Xiaolong Bai, Xueqiang Wang, and Xiaofeng Wang.
\newblock Privacy loss in apple's implementation of differential privacy on
  macos 10.12.
\newblock {\em arXiv preprint arXiv:1709.02753}, 2017.

\bibitem{apple2017dp}
Differential~Privacy Team.
\newblock Learning with privacy at scale.
\newblock Apple, 2017.

\bibitem{twitter_incq3}
Inc. Twitter.
\newblock Twitter q3 2021 earnings report, 2021.

\bibitem{winer2009new}
Russell~S Winer.
\newblock New communications approaches in marketing: Issues and research
  directions.
\newblock {\em Journal of interactive marketing}, 23(2):108--117, 2009.

\end{thebibliography}
\section*{Appendix}
\begin{proof}[Proof of Theorem~\ref{th:revenue}]
	We show that $g$ in Equation~\ref{eq:sol} is the best approximation (i.e., expected value) of the unknown $y_\alpha^t$ based on $\tilde{U}$. It minimizes
	\begin{equation}
		\label{eq:together}
		\sum_\alpha\left(\sum_v(x_{v,\alpha}\cdot\bar{r}_v^t)-y_\alpha^t\right)^2
	\end{equation}
	The revenue $y_\alpha^t$ is the sum of revenues of the users with $\alpha_i=\alpha$.
	Moreover, the users $U$ can be divided into disjoint sets based on the conversion values (similarly how $\tilde{U}_{v}$  is defined), expressing $y_\alpha^t$ with a double summation.
	\begin{equation*}
		y_\alpha^t=\sum_{i=1}^{|U|}\mathbbm{1}(\alpha_i=\alpha)\cdot r_i^t=
		\sum_v\sum_{U_v}\mathbbm{1}(\alpha_i=\alpha)\cdot r_i^t
	\end{equation*}
	Combining this with Equation~\ref{eq:together} we get
	\begin{align*}
		\sum_\alpha\left(\sum_vx_{v,\alpha}\cdot\bar{r}_v^t-\sum_v\sum_{U_v}\mathbbm{1}(\alpha_i=\alpha)\cdot r_i^t\right)^2= \\
		\sum_\alpha\left(\sum_v\left(x_{v,\alpha}\cdot\bar{r}_v^t-\sum_{U_v}\mathbbm{1}(\alpha_i=\alpha)\cdot r_i^t\right)\right)^2
	\end{align*}
	The above equation is minimal if the absolute elements of the summation over $\alpha$ are minimal, hence, it is enough to show this minimality for an arbitrary $\alpha$.
	It is trivial that $x_{v,\alpha}=\sum_{U_{v}}\mathbbm{1}(\alpha_i=\alpha)$.
	On the other hand, when only $\tilde{U}_{v}$ is available (instead of $U_{v}$) then for any specific conversion value $v$ it is unknown which particular users $i$ is counted in $x_{v,\alpha}$.
	Consequently, instead of the unknown $\alpha_i$ we use a random variable noted as $\check{\alpha}_i$:
	user $i$'s origin within $\tilde{U}_v$ (i.e., with conversion value $v$) is $\alpha$ with probability $\frac{x_{v,\alpha}}{|\tilde{U}_v|}$.
	Finally, we show that the expected value of these probabilities is $g$ as defined in Equation~\ref{eq:sol}, hence the expected error is zero.
	\begin{align*}
		\sum_{U_v}\mathbbm{1}(\alpha_i=\alpha)\cdot r_i^t=
		\sum_{\tilde{U}_v}\mathbb{E}[\mathbbm{1}(\check{\alpha}_i=\alpha)]\cdot r_i^t= \\
		\sum_{\tilde{U}_v}\frac{x_{v,\alpha}}{|\tilde{U}_v|}\cdot r_i^t=
		x_{v,\alpha}\cdot\sum_{\tilde{U}_v}\frac{r_i^t}{|\tilde{U}_v|}=
		x_{v,\alpha}\cdot\bar{r}_v^t
	\end{align*}
	\hfill\qedhere
\end{proof}
\begin{proof}[Proof of Theorem~\ref{th:revenueP}]
	The proof is similar to the case when $p<2$ with a minor modification, as Equation~\ref{eq:sol} does not consider \texttt{null} values.
	It is trivial that the revenue of $\alpha$ is under approximated if all buckets with value \texttt{null} are treated as zero.
	In other words, Equation~\ref{eq:sol} does not attribute the leftover revenue to the campaigns.
	In case $\hat{x}_{v,\alpha}=\texttt{null}$, the original value could be anything between $0\le x_{v,\alpha}\le \sum_U\mathbbm{1}(v_i=v)\le p-1$.
	To take this into account, the value of $x_{v,\alpha}$ can be seen as a random variable denoted as $\check{x}_{v,\alpha}$.
	For a specific $v$ the distribution over $\alpha$'s itself is not as easy to formalize as in the case of $\check{\alpha}_i$,
	because while those are I.I.D. variables, $\check{x}_{v,\alpha}$'s are not independent (as their sum must be equal with $\sum_U\mathbbm{1}(v_i=v)$).
	In this case, for a specific conversion value $v$ we only need the expected value,
	which is some probability ($\fbox{$\phantom{12}$}$) multiplied with the user count as shown in Equation \ref{eq:rev_all}.
	The exact probability must be between the values defined by \textbf{U} and \textbf{N},
	as they capture the two extreme cases depending on the amount of useful information contained in the \texttt{null} bucket.
	The first is when the information in the \texttt{null} bucket encapsulates the $v$'s distribution perfectly.
	This is the case, for example, when $v$ is the only conversion value with a user count below $p$, in this case,
	\textbf{N} gives the best approximation of $v$'s distribution because they are identical.
	The second is when the information within the \texttt{null} bucket is useless.
	For instance, all conversion values' user counts are below $p$, and their corresponding distributions are completely different from each other.
	In this scenario \textbf{U} corresponds to the best unbiased approximation (i.e., uniform random guess).
	Hence, any information within the \texttt{null} bucket could be covered with a convex combination of \textbf{U} and \textbf{N}.
	By multiplying this expected probability with $\bar{r}_v^t$ the rest of the proof follows the proof of Theorem \ref{th:revenue}.
	\hfill$\qedhere$
\end{proof}
\end{document}